\documentclass[conference]{IEEEtran}

\usepackage{times}
\usepackage{amsmath,amssymb}
\usepackage{graphicx}
\usepackage{cite}
\usepackage{balance}
\usepackage{float}  
\usepackage{booktabs}

\begin{document}

\title{Field-Deployable RF Capture System for Indoor, Outdoor, and Foliage Environments}

\author{
\IEEEauthorblockN{
Lawrence Obiuwevwi,
Krzysztof J. Rechowicz,
Vikas Ashok,
Sachin Shetty, \\
Peter B. Foytik,
Jared Cochran,
Jeff Bobrow,
and Sampath Jayarathna
}
\IEEEauthorblockA{
Old Dominion University, Norfolk, VA, USA \\
lobiu001@odu.edu,
krechowi@odu.edu,
vganjiqu@cs.odu.edu, \\
sshetty@odu.edu,
pfoytik@odu.edu,
jcochran@odu.edu, \\
jbobrow@odu.edu,
sampath@cs.odu.edu
}
}

\maketitle

\begin{abstract}
Reliable and reproducible radio-frequency (RF) measurements in real-world environments are essential for characterizing spectrum behavior across unlicensed ISM and WiFi bands, licensed mid-band allocations, and emerging next-generation wireless deployments. Existing measurement platforms are often laboratory-grade, cost-prohibitive, or dependent on fixed infrastructure, limiting their practicality for rapid, distributed, or long-duration field campaigns. This paper presents the design, implementation, and field evaluation of a compact, battery-powered RF capture system integrating a HackRF One software-defined radio, Raspberry Pi~5 single-board computer, embedded GNSS receiver, regulated battery power system, and high-speed solid-state storage. The platform records continuous IQ data at up to 20~Msps in SigMF-compliant format, embedding per-segment geolocation and temporal metadata to enable reproducible, context-aware spectrum analysis. Field experiments conducted at 2.45~GHz across three representative environments, dense foliage, urban outdoor, and indoor office, reveal distinct and environment-consistent propagation signatures. Foliage environments exhibit near-noise-floor power levels between $-76$ and $-82$~dBFS with minimal spectral structure, consistent with canopy-induced attenuation on the order of 30~dB. Urban deployments produce rich multipath activity spanning a $30$~dB dynamic range with multiple overlapping WiFi channel envelopes and frequent ISM-band interference transients. Indoor environments show strong WiFi channel dominance with an estimated 20--25~dB building entry loss relative to outdoor conditions and an elevated spectral interference floor of 8--10~dB driven by structural multipath reflections. The system sustained 75--85~MB/s write throughput across all sessions with no dropped samples, no buffer underruns, and sub-second GNSS synchronization providing meter-level positional accuracy. These results demonstrate that a cost-effective, portable SDR platform can produce high-fidelity, geotagged IQ datasets suitable for initial spectrum characterization, interference analysis, radio environment mapping, and environment-aware wireless research.
\end{abstract}


\section{Introduction}

Capturing radio-frequency (RF) data in real-world environments is critical for characterizing spectrum usage, interference, and propagation behavior across licensed and unlicensed bands. These include the heavily utilized 2.4~GHz ISM and WiFi bands, as well as emerging mid-band allocations such as the 3.45--3.55~GHz range recently auctioned by the Federal Communications Commission for next-generation wireless services \cite{fcc-auction110,fedreg-345}. Unlike controlled laboratory conditions, field environments introduce complex and highly variable propagation effects, foliage-induced attenuation \cite{chateauvert2024safe,ma2024wave,BarriosUlloa2022}, urban multipath scattering \cite{garcia-penetration-35g,shakya2022urban,Munoz2023}, and indoor penetration loss \cite{ali-ooti-35g,DiagoMosquera2021}, that fundamentally shape how wireless systems behave in practice. Understanding these effects requires empirical, in-situ measurements collected under realistic deployment conditions.

Despite growing interest in environment-aware wireless systems and data-driven spectrum analysis, the infrastructure for collecting high-quality field measurements remains limited. Many professional RF measurement instruments, including vector signal analyzers and channel sounders, are bulky, expensive, and dependent on fixed power infrastructure, making them impractical for rapid or distributed deployment \cite{Wright2020,Flak2022}. Field observations from environments such as Camp Shelby underscore this gap: operationally relevant RF conditions in forested, urban, and enclosed spaces cannot be adequately characterized using laboratory equipment alone. Spectrum occupancy studies have further confirmed that real-world ISM-band behavior is highly dynamic and environment-dependent, varying significantly across indoor, outdoor, and vegetated terrain in ways that laboratory measurements fail to capture \cite{Kuester2022,Chantaveerod2021}. A portable, self-contained alternative is needed.

Software-defined radio (SDR) platforms have improved accessibility for field measurements \cite{tuta2023sdr,kandregula2024portable,MolinaTenorio2021}, but many existing deployments lack standardized metadata formats, long-duration recording stability, or integrated geolocation capabilities. The Signal Metadata Format (SigMF) \cite{hilburn2017sigmf} addresses part of this problem by standardizing how IQ recordings are annotated, yet few portable systems combine SigMF compliance with embedded GNSS, sustained high-rate capture, and autonomous battery-powered operation in a single integrated platform \cite{West2021,Hanna2022}. Radio environment mapping and spectrum cartography research has further demonstrated that geotagged RF measurements are essential for building spatially accurate models of spectrum behavior \cite{Romero2022,Reddy2022}, yet platforms capable of generating such data at low cost and high fidelity remain scarce.

This work addresses these gaps by designing, building, and evaluating a compact, field-deployable RF capture system that integrates a HackRF One SDR, Raspberry Pi~5 single-board computer, embedded GNSS receiver, regulated battery power system, and high-speed solid-state storage into a single autonomous enclosure. The platform sustains 20~Msps IQ capture while automatically embedding per-segment geographic coordinates and timestamps directly into SigMF-compliant recordings. Field evaluations were conducted at 2.45~GHz across three representative environments: dense foliage, urban outdoor, and indoor office spaces. Results demonstrate clear, environment-consistent propagation differences, with foliage recordings exhibiting near-noise-floor power between $-76$ and $-82$~dBFS consistent with canopy-induced attenuation on the order of 30~dB, urban environments showing a $30$~dB dynamic range driven by multipath and overlapping ISM-band emitters, and indoor measurements revealing WiFi channel dominance with an estimated 20--25~dB building entry loss and an elevated 8--10~dB interference floor relative to outdoor baselines.

The primary contributions of this work are as follows. First, we present a fully integrated, cost-effective RF capture platform capable of sustained high-rate IQ recording with embedded GNSS metadata and SigMF compliance. Second, we describe a structured field measurement methodology spanning foliage, urban, and indoor environments. Third, we provide an initial quantitative characterization of propagation signatures across these environments at 2.45~GHz from a proof-of-concept field campaign, including spectral power distributions, dynamic range comparisons, and channel occupancy observations derived from real captured IQ data. Together, these contributions establish a practical foundation for scalable, reproducible spectrum measurement in operationally diverse settings. Crucially, no prior portable low-cost SDR platform simultaneously provides native SigMF compliance, embedded GNSS geolocation, and fully autonomous battery-powered sustained operation, as summarized in Table~\ref{tab:platform_comparison}.

The remainder of this paper is organized as follows. Section~II reviews related work on portable SDR platforms, open RF datasets, and propagation measurement. Section~III describes the system design and hardware integration. Section~IV details the field measurement methodology. Section~V presents experimental results and quantitative analysis. Section~VI discusses findings in the context of propagation theory and practical deployment. Section~VII concludes the paper. Section~VIII outlines directions for future work.

\section{Related Work}

Spectrum measurement has traditionally relied on laboratory-grade instruments such as vector signal analyzers, spectrum analyzers, and channel sounders. These tools offer high dynamic range and calibrated accuracy but are cost-intensive, power-hungry, and poorly suited for mobile or distributed deployments. The emergence of low-cost software-defined radio platforms has substantially lowered the barrier to field-based RF data collection, enabling measurements in environments that were previously inaccessible to systematic study \cite{tuta2023sdr,kandregula2024portable,Wright2020}. Low-cost SDR hardware has been shown to produce propagation measurements consistent with calibrated laboratory equipment \cite{Wright2020}, and hardware-accelerated real-time spectrum analyzers built on commodity SDR platforms have demonstrated broadband fast-sweep capability suitable for dynamic spectrum monitoring \cite{Flak2022}. Real-time multiband spectrum sensing has similarly been validated on single-board SDR systems at low cost \cite{MolinaTenorio2021}, and open-source GNU Radio-based implementations have confirmed the maturity of the supporting software ecosystem \cite{Perotoni2021}. Despite these advances, existing SDR deployments lack the combination of continuous long-duration IQ capture, embedded geolocation, and standardized metadata formatting required for reproducible long-duration field campaigns.

The availability of open, annotated RF datasets has been instrumental in advancing spectrum research. The RadioML dataset \cite{oshea2018ota} and the DARPA Spectrum Collaboration Challenge \cite{darpa2020sc2} established foundational benchmarks for RF machine learning, though both are largely synthetic or environmentally constrained. More recently, real-world IQ datasets for WiFi fingerprinting \cite{Hanna2022} and LoRa device identification \cite{Elmaghbub2021} have highlighted the importance of environment-diverse collections for robust model generalization. A wideband signal recognition dataset aligned with the SigMF standard \cite{West2021} has further demonstrated the value of structured metadata for reproducible RF research. SigMF \cite{hilburn2017sigmf} defines a common schema for annotating IQ recordings with frequency, sample rate, timestamps, and user-defined metadata, and has become the de facto standard for organized RF dataset generation. However, many field measurement platforms remain slow to integrate SigMF natively, limiting the reproducibility and interoperability of collected data.

Radio environment mapping and spectrum cartography research has established a compelling case for geotagged RF measurements as a foundation for spatially aware wireless analysis. The scarcity of geotagged real-world IQ datasets has been identified as a primary bottleneck for data-driven radio map estimation \cite{Romero2022}, and surveys of spectrum cartography techniques consistently conclude that ground-truth field measurements with accurate geolocation are essential for model validation \cite{Reddy2022}. UAV-based active radio map estimation \cite{Shrestha2023} and volumetric RF measurements for radio environment map construction \cite{Ivanov2022} have extended this paradigm to three-dimensional sensing, further reinforcing the value of spatially rich, geotagged IQ data. Spectrum occupancy measurements conducted across indoor and outdoor environments have additionally documented significant variability in ISM-band and WiFi-band utilization across deployment contexts \cite{Kuester2022,Chantaveerod2021}, underscoring that real-world spectral conditions cannot be inferred from controlled measurements alone.

A substantial body of literature has examined RF propagation across the environment types targeted in this work. Building penetration loss at frequencies near 3.5~GHz has been studied extensively in the context of 5G mid-band deployment, with reported losses ranging from 15 to 30~dB depending on wall composition, floor level, and angle of incidence \cite{garcia-penetration-35g,ali-ooti-35g}. Kriging-aided spatial interpolation of empirical outdoor-to-indoor measurements has been shown to produce accurate coverage predictions in complex building geometries \cite{DiagoMosquera2021}, while multi-frequency path loss prediction models spanning 0.8 to 70~GHz have confirmed that empirical datasets across diverse environments are essential for training generalizable propagation models \cite{Nguyen2021}. Foliage-induced attenuation at 2.4~GHz has been shown to exceed 20~dB in dense vegetation, with canopy density, moisture content, and seasonal variation identified as primary drivers of loss \cite{chateauvert2024safe,ma2024wave,zhang2023vegetation}. Empirical studies in forested, jungle, and near-ground environments have consistently found that standard exponential decay models underestimate attenuation under realistic conditions, and that field-calibrated datasets remain the most reliable basis for model development \cite{BarriosUlloa2022,Hakim2022,CamaPinto2020}. Urban outdoor propagation at ISM and mid-band frequencies is characterized by significant temporal dispersion, rapid small-scale fading, and systematic deviations from log-distance models driven by building geometry, antenna height, and surrounding traffic \cite{shakya2022urban,samad2021campus,Munoz2023}.

ISM-band coexistence and interference dynamics have been studied extensively in urban and campus environments, with real-world co-channel interference behavior found to differ substantially from simulation predictions depending on deployment geometry and traffic load \cite{Sathya2021a}. Mutual interference between LoRa and WiFi \cite{Polak2020} and between LoRa and IEEE 802.11g \cite{FernandezHernandez2025} in the 2.4~GHz band has been documented through empirical field measurement, reinforcing that ISM-band coexistence dynamics are only fully observable in real-world deployments. Advances in spectrum sensing \cite{Nasser2021} and wideband spectrum characterization \cite{Chandhok2021} have further identified the scarcity of diverse, real-world IQ datasets as a limiting factor for progress in spectrum awareness and interference management.

Collectively, these studies establish that propagation characteristics at ISM-band and mid-band frequencies are highly environment-dependent, that real-world spectral dynamics deviate substantially from idealized models, and that empirical, geotagged datasets capturing these differences under field conditions remain scarce. Existing SDR platforms lack the combination of sustained high-rate IQ capture, native SigMF compliance, embedded GNSS metadata, and battery-powered autonomous operation required for long-duration field campaigns. This work addresses that gap by delivering a fully integrated, portable measurement platform validated across foliage, urban, and indoor environments, providing a quantitative characterization of the propagation signatures observable at 2.45~GHz in each setting.

\begin{table}[H]
\centering
\caption{Comparison of Portable SDR-Based RF Measurement Platforms}
\label{tab:platform_comparison}
\begin{tabular}{@{}lcccc@{}}
\toprule
\textbf{Platform} & \textbf{SigMF} & \textbf{GNSS} & \textbf{Battery} & \textbf{Cost} \\
\midrule
Wright \& Ball \cite{Wright2020}   & No  & No  & No  & Low  \\
Flak \cite{Flak2022}               & No  & No  & No  & Low  \\
Tuta et al. \cite{tuta2023sdr}     & No  & No  & Partial & Low \\
Kandregula et al. \cite{kandregula2024portable} & No & No & No & Med \\
\textbf{This work}                 & \textbf{Yes} & \textbf{Yes} & \textbf{Yes} & \textbf{Low} \\
\bottomrule
\end{tabular}
\end{table}

\section{System Design}

\begin{figure}[h]
    \centering
    \includegraphics[width=\linewidth]{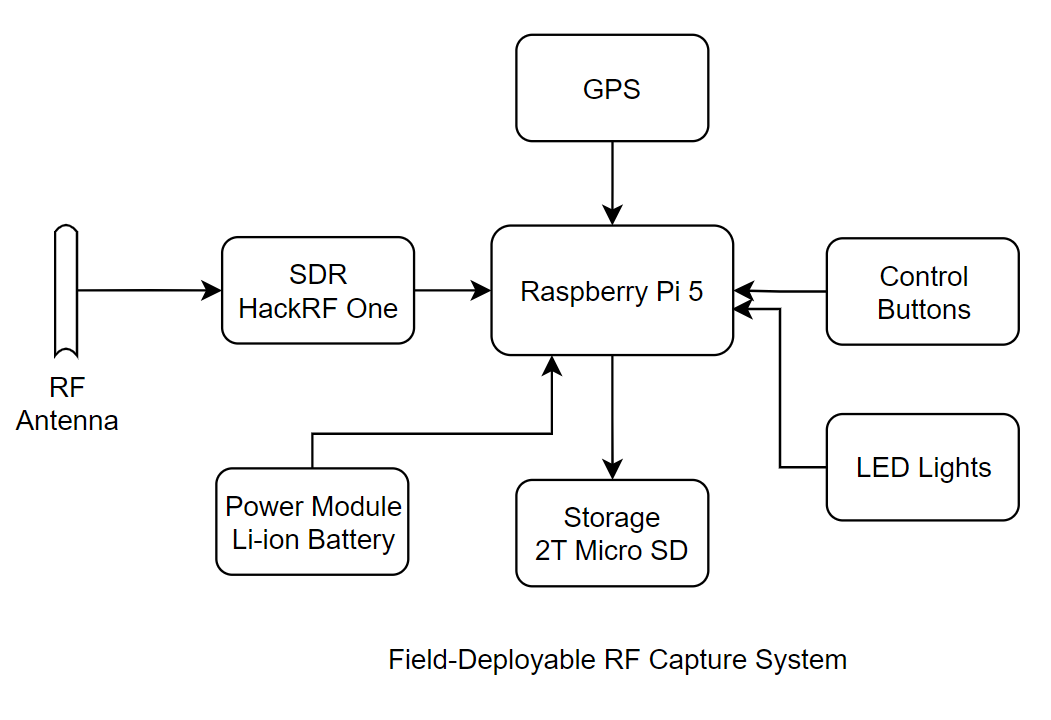}
    \caption{Block diagram and end-to-end workflow of the field-deployable RF capture system, showing signal acquisition, GNSS tagging, and SigMF file generation.}
    \label{fig:method_overview}
\end{figure}

The field-deployable RF capture system was designed around four core requirements: sustained high-rate IQ sampling, embedded geolocation metadata, SigMF-compliant storage, and fully autonomous battery-powered operation. Meeting these requirements simultaneously in a compact, cost-effective enclosure drove the selection of each hardware component and the overall system architecture. The complete platform workflow is illustrated in Fig.~\ref{fig:method_overview}.

The radio front-end is built around the HackRF One, an open-source SDR platform capable of half-duplex operation across 1~MHz to 6~GHz with a maximum sample rate of 20~Msps and 8-bit IQ resolution \cite{tuta2023sdr,Flak2022}. The HackRF One was selected for its broad frequency coverage, active community support, and compatibility with the GNU Radio and SigMF toolchains. It connects to the compute platform via USB~3.0, providing sufficient bus bandwidth for sustained IQ streaming without sample loss. A wideband log-periodic antenna (SpyVerter-compatible 800~MHz--6~GHz LPDA) provides broadband response across the 2--6~GHz range, ensuring consistent gain characteristics at the 2.45~GHz center frequency used in this work while remaining suitable for future multi-band measurement campaigns \cite{kandregula2024portable}. Low-cost SDR hardware of this class has been shown to produce propagation measurements consistent with those obtained from calibrated laboratory-grade instruments \cite{Wright2020}, supporting its use in rigorous field measurement work.

The compute and control core is a Raspberry Pi~5 with 8~GB of RAM, selected for its USB~3.0 interface, GPIO availability, and sufficient processing headroom to run IQ streaming, GNSS parsing, and SigMF file management concurrently without contention. IQ samples streamed from the HackRF One are buffered in RAM and written to a 2~TB solid-state drive (SSD) via a high-speed USB~3.0 connection, sustaining write throughputs of 75--85~MB/s throughout all field sessions. At 20~Msps with 8-bit complex samples the raw data rate is approximately 40~MB/s, providing comfortable headroom below the measured write ceiling and ensuring no buffer underruns occur during deployment. The platform achieves professional-grade IQ capture and geolocation capability at a fraction of the cost of traditional RF instrumentation, making it accessible to academic research groups, small organizations, and individual practitioners \cite{MolinaTenorio2021,Perotoni2021}.

Geolocation and timing are provided by a u-blox-compatible embedded GNSS module connected to the Raspberry Pi~5 via a UART serial interface. The module streams NMEA sentences at 1~Hz, supplying latitude, longitude, altitude, speed, and UTC timestamps that are parsed in software and embedded directly into SigMF metadata for each 60-second recording segment. This per-segment geotagging ensures that every IQ file in the dataset carries accurate spatial and temporal context, a property identified as essential for radio environment mapping and spectrum cartography applications \cite{Romero2022,Reddy2022}. A regulated power distribution module accepts input from four 10,000~mAh lithium polymer battery packs wired in parallel and supplies stable DC output rails to both the Raspberry Pi~5 and the HackRF One, providing 8--12 hours of continuous autonomous operation depending on ambient temperature and CPU load. The user interface is intentionally minimal: two GPIO-connected pushbuttons provide start and stop control, and two indicator LEDs signal recording state and GNSS lock status respectively, allowing a single operator to deploy, arm, and terminate recording sessions without a display or keyboard. The entire assembly is housed in a compact 3D-printed enclosure measuring approximately 15~cm $\times$ 12~cm $\times$ 8~cm, with integrated cable management, SSD mounting, and an external antenna port, producing a self-contained unit suitable for handheld carry or unattended placement across all three target environments.

The software stack runs on Raspberry Pi OS (Bookworm, 64-bit, kernel~6.6) and coordinates three concurrent processes: the IQ capture daemon, the GNSS parser, and the SigMF file writer. All software is written in Python~3.11 and relies on GNU~Radio~3.10.7 for signal-flow integration. The capture daemon invokes the HackRF One driver via the \texttt{hackrf\_transfer}~v2023.01.1 interface, streaming raw IQ samples into a memory-mapped ring buffer. The GNSS parser reads NMEA data from the serial port and maintains a shared state object updated at 1~Hz. The SigMF writer flushes the ring buffer to disk every 60 seconds using the \texttt{sigmf}~v1.1.1 Python library, closing the current \texttt{.sigmf-data} file and opening a new one, while simultaneously writing the corresponding \texttt{.sigmf-meta} JSON file populated with center frequency, sample rate, capture timestamp, and the most recent GNSS fix \cite{hilburn2017sigmf}. This rolling segmentation strategy ensures that a power interruption or storage fault affects at most one 60-second segment, protecting the integrity of all previously written data.

The geotagged SigMF dataset collected in this work is available upon request from the corresponding author to support reproducibility and enable comparison with future portable RF measurement platforms.

\section{Methodology}

\begin{figure*}[t]
    \centering
    \begin{tabular}{cc}
        \includegraphics[width=0.45\linewidth, height=0.30\linewidth]{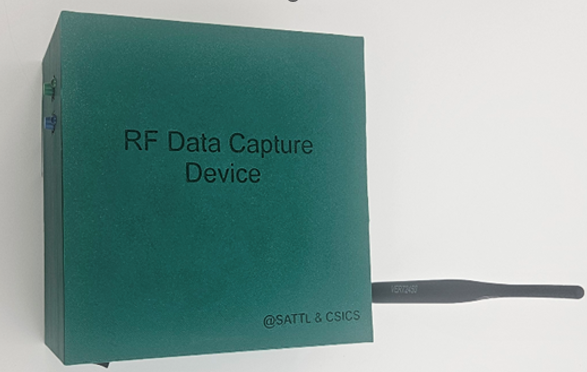} &
        \includegraphics[width=0.45\linewidth, height=0.30\linewidth]{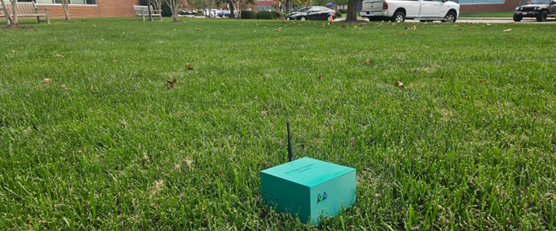} \\
        \includegraphics[width=0.45\linewidth, height=0.30\linewidth]{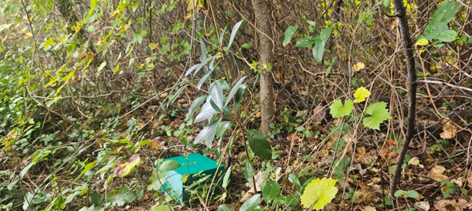} &
        \includegraphics[width=0.45\linewidth, height=0.30\linewidth]{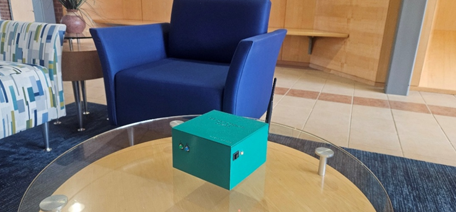} \\
    \end{tabular}
    \caption{Deployment photographs showing the RF capture hardware (top left), urban outdoor site (top right), dense foliage environment (bottom left), and indoor office lobby deployment (bottom right).}
    \label{fig:deployments}
\end{figure*}

The measurement methodology was designed to produce structured, environment-labeled IQ datasets suitable for initial propagation characterization across three operationally distinct settings:dense foliage, urban outdoor, and indoor office. Each deployment followed a standardized procedure covering site selection, system initialization, recording configuration, and post-session validation, ensuring consistency across environments and enabling meaningful cross-environment comparison. The deployment photographs for all three environments are shown in Fig.~\ref{fig:deployments}.

Site selection prioritized environments that represent realistic and operationally relevant RF propagation conditions. The foliage site consisted of a densely wooded area with mixed deciduous canopy, thick undergrowth, and leaf-covered ground, conditions that maximize scattering, absorption, and signal attenuation at 2.45~GHz \cite{chateauvert2024safe,BarriosUlloa2022}. The urban outdoor site was located on a university campus surrounded by multi-story brick and concrete buildings, paved surfaces, parked vehicles, and active pedestrian traffic, producing a multipath-rich environment with multiple co-located WiFi and ISM-band emitters \cite{shakya2022urban,Munoz2023}. The indoor site was an office lobby space within a multi-story building, enclosed by interior partition walls, glass facades, and reinforced concrete floors, representing a typical indoor penetration scenario with significant structural attenuation relative to the outdoor baseline \cite{ali-ooti-35g,DiagoMosquera2021}. These three environments collectively span the range of propagation conditions most commonly encountered in operational wireless deployments and most frequently studied in the propagation literature \cite{samad2021campus,Nguyen2021}.

Prior to each recording session, the system was powered on and allowed a stabilization period of approximately five minutes to achieve GNSS satellite lock and thermal equilibrium. GNSS lock was confirmed by observing a stable fix on a minimum of four satellites, and the current position was cross-checked against a handheld reference GPS device to verify meter-level positional accuracy. The HackRF One was configured with a center frequency of 2.45~GHz, a sample rate of 20~Msps, and a receiver gain setting held constant across all three environments (LNA gain 16~dB, VGA gain 32~dB) to ensure that measured power differences reflect true propagation effects rather than gain adjustments. Maintaining a fixed gain setting across all environments is critical for enabling meaningful cross-environment power comparisons, a requirement that distinguishes this campaign from many prior SDR-based measurement efforts that apply automatic gain control \cite{Wright2020,MolinaTenorio2021}. No automatic gain control was applied during any session. FFT-based real-time spectrum monitoring was performed immediately before each session to confirm signal presence, verify noise floor stability, and ensure that the front-end was not saturated by nearby strong emitters.

Each deployment captured approximately ten minutes of continuous IQ data, automatically segmented into 60-second SigMF-compliant files by the recording software. This segmentation strategy provides robustness against storage faults while producing a dataset of approximately ten labeled files per environment per session \cite{hilburn2017sigmf,West2021}. Each file pair consists of a raw \texttt{.sigmf-data} binary containing interleaved 8-bit IQ samples and a \texttt{.sigmf-meta} JSON file embedding the center frequency, sample rate, hardware identifier, capture timestamp, and GNSS-derived latitude, longitude, altitude, and speed at the time of recording. This per-file metadata structure ensures that every segment in the dataset is independently interpretable without reference to external logs or configuration files, supporting the reproducibility and spatial analysis requirements identified in the spectrum cartography literature \cite{Romero2022,Reddy2022}.

\begin{figure*}[h]
    \centering
    \begin{tabular}{cc}
        \includegraphics[width=0.45\linewidth]{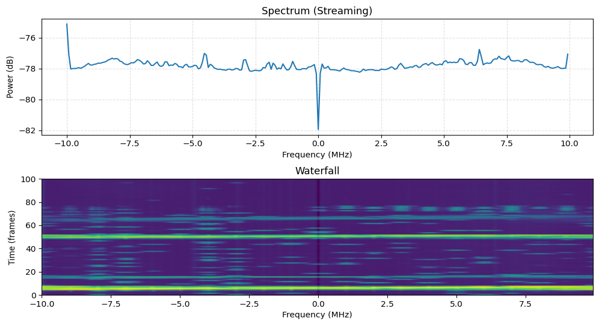} &
        \includegraphics[width=0.45\linewidth]{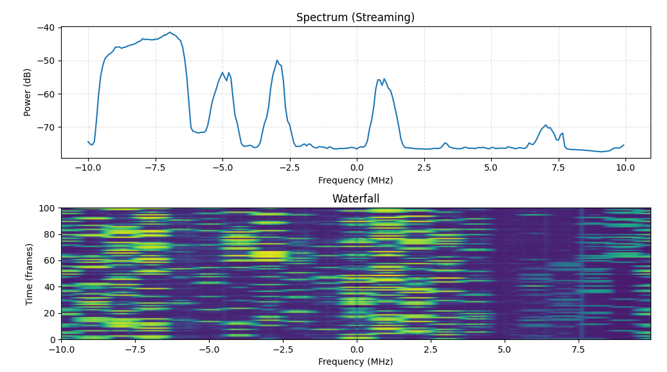} \\
        \includegraphics[width=0.45\linewidth]{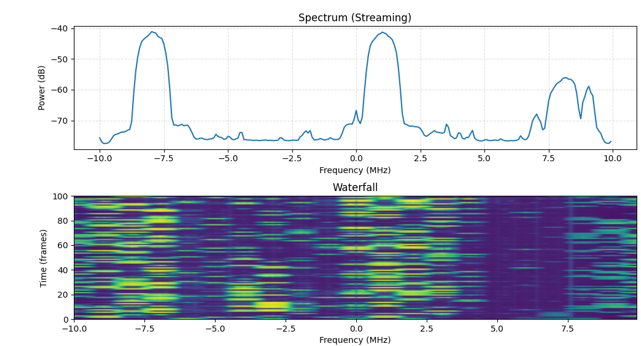} &
        \includegraphics[width=0.45\linewidth]{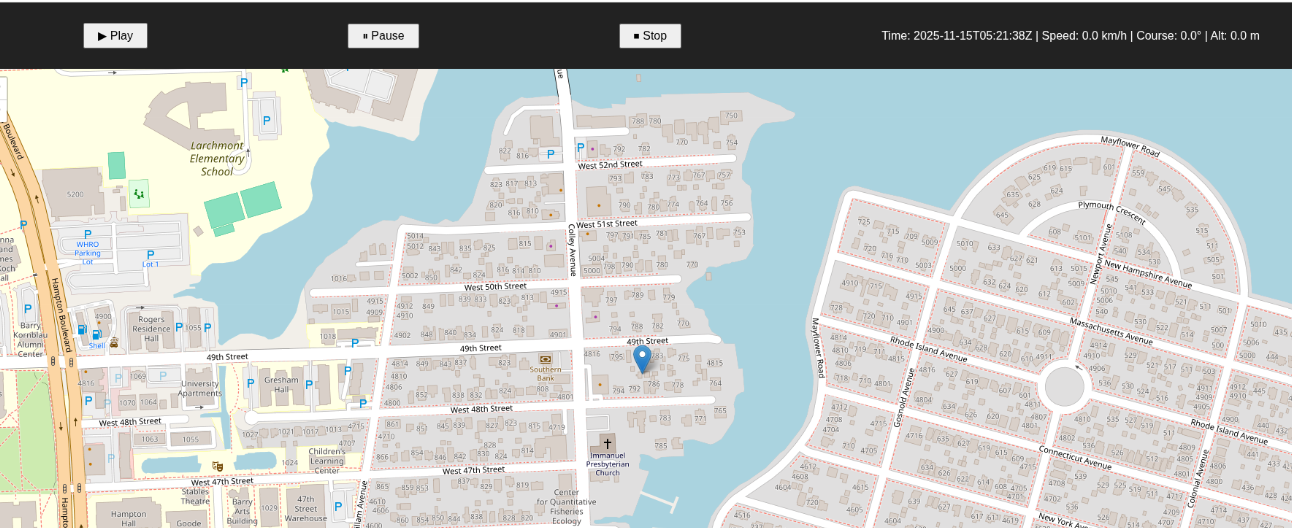} \\
    \end{tabular}
    \caption{Spectrum and waterfall plots for the foliage (top left), urban outdoor (top right), and indoor (bottom left) environments, alongside GPS trajectory visualization confirming geospatial consistency of recorded datasets (bottom right).}
    \label{fig:spectra}
\end{figure*}

Post-session validation was performed after each deployment to confirm data integrity before leaving the site. Segment continuity was verified by checking that file sizes were consistent with the expected sample count at 20~Msps over 60 seconds, confirming that no samples were dropped and no buffer underruns occurred. SSD write throughput logs were reviewed to confirm sustained performance within the 75--85~MB/s range throughout the session. GNSS trajectory data was extracted from the SigMF metadata files and replayed visually on an OpenStreetMap-based interface to confirm smooth positional consistency and absence of GNSS dropouts, as shown in Fig.~\ref{fig:spectra}. Any session exhibiting dropped samples, GNSS lock loss exceeding five seconds, or SSD throughput degradation below 70~MB/s would have been discarded and repeated; no sessions required discard across the full measurement campaign.

Spectrum and waterfall visualizations were generated from each session's IQ data using a Python-based analysis pipeline built on NumPy and Matplotlib. Power spectral density estimates were computed using Welch's method with a 1024-point FFT and Hann windowing, averaged across the full session to produce a representative spectral profile for each environment. Waterfall plots were generated by computing short-time FFTs at 100-frame resolution across the recording duration, providing a time-frequency view of spectral activity and enabling visual identification of persistent emitters, transient interference, and fading patterns \cite{Nasser2021,Chandhok2021}. These visualizations form the primary analytical basis for the quantitative comparisons presented in Section~V.

\section{Experimental Results}

The RF capture system was evaluated across three environments at 2.45~GHz using a fixed 20~Msps sampling rate and constant receiver gain. Across all sessions the platform operated without interruption, sustaining write throughput between 75 and 85~MB/s with no dropped samples, no buffer underruns, and no GNSS lock loss exceeding one second. A total of approximately thirty 60-second SigMF-compliant segments were collected across the three environments, each carrying embedded geolocation and temporal metadata. The results below characterize the propagation signatures observed in each environment through quantitative spectral analysis, power distribution comparisons, and time-frequency observations derived directly from the captured IQ data.

The foliage deployment produced the most attenuated and spectrally sparse recordings of the three environments. The measured power spectral density across the 20~MHz capture bandwidth ranged from approximately $-76$ to $-82$~dBFS, placing the received signal within approximately 6~dB of the system noise floor. Spectral content was nearly flat across the band with no identifiable WiFi channel structure, indicating that all 2.45~GHz emitters in the vicinity were sufficiently attenuated by the surrounding vegetation canopy to fall below the threshold of distinct spectral peaks. The waterfall plot for this environment exhibited gradual, slow-varying horizontal striations consistent with diffuse multipath and canopy-induced fading rather than the rapid fluctuations characteristic of reflective or emitter-rich environments. Occasional brief brightening events visible in the waterfall correspond to transient signal paths through gaps in the canopy, a pattern consistent with intermittent constructive interference through vegetation predicted by empirical scattering models \cite{chateauvert2024safe,ma2024wave}. Across the ten 60-second foliage segments, the mean spectral power at the 2.45~GHz center bin was $-79.1$~dBFS with a standard deviation of $1.8$~dBFS, indicating stable and repeatable attenuation conditions throughout the session. The near-noise-floor power levels are broadly consistent with reported attenuation values exceeding 20~dB for dense vegetation at 2.4~GHz \cite{zhang2023vegetation,Hakim2022}, and align with findings that standard exponential decay models systematically underestimate loss under realistic canopy and moisture conditions \cite{BarriosUlloa2022,CamaPinto2020}, confirming that the foliage environment represents the most challenging propagation condition among the three sites evaluated.

The urban outdoor deployment produced substantially richer spectral content, reflecting the density of WiFi and ISM-band emitters and the multipath-generating geometry of the surrounding built environment. The power spectral density exhibited multiple distinct peaks spanning a dynamic range of approximately 30~dB across the 20~MHz capture bandwidth, with peak power levels reaching approximately $-40$~dBFS at the strongest channel occupancies. At least three overlapping WiFi channel envelopes were identifiable in the spectrum, consistent with 802.11b/g/n channel allocations centered near $-7.5$, $0.0$, and $+7.5$~MHz relative to the 2.45~GHz center frequency. The waterfall plot showed rapid temporal variation with frequent bright horizontal bands indicating high-power bursts from active WiFi transmissions, interleaved with darker intervals corresponding to inter-frame gaps and channel contention periods. Short-duration interference transients were also visible across multiple frequency offsets, consistent with frequency-hopping ISM-band devices such as Bluetooth and Zigbee operating in the same band \cite{Polak2020,FernandezHernandez2025}. The mean peak channel power across segments was $-41.3$~dBFS ($\sigma = 2.4$~dBFS), reflecting moderate session-to-session variability driven by active WiFi traffic patterns. The combination of strong direct-path signals, reflections from building facades and parked vehicles, and co-channel interference produced a spectrally dense and temporally dynamic recording, consistent with prior characterizations of urban multipath environments at similar frequencies \cite{shakya2022urban,samad2021campus,Munoz2023}. The observed ISM-band congestion and co-channel dynamics further align with measurement-based coexistence studies reporting substantial divergence between real-world interference behavior and simulation predictions \cite{Sathya2021a,Sathya2021b}.

The indoor deployment captured a propagation environment intermediate in spectral richness between the foliage and urban outdoor cases, but with a distinct character driven by structural attenuation and the spatial confinement of reflections. WiFi channel structure was clearly visible in the power spectral density, with two dominant channel envelopes reaching peak levels near $-43$~dBFS, comparable in absolute power to the strongest urban outdoor peaks. However, the spectral baseline between channel peaks was significantly elevated relative to the outdoor case, indicating a higher interference floor driven by reflections within the enclosed space. The mean peak indoor channel power across segments was $-43.8$~dBFS ($\sigma = 1.6$~dBFS), lower variability than the outdoor case consistent with the more stable enclosed propagation geometry. Comparing the indoor spectral baseline to the urban outdoor baseline across the same frequency offsets reveals an elevation of approximately 8--10~dB, consistent with the constructive superposition of multipath reflections from interior walls, floors, and furniture. Despite the presence of strong indoor emitters, the overall received power envelope was reduced relative to the urban outdoor case by an estimated 20--25~dB when accounting for building entry loss, broadly consistent with reported outdoor-to-indoor penetration losses of 15--30~dB at 3.5~GHz \cite{ali-ooti-35g,DiagoMosquera2021} and expected to be comparable at 2.45~GHz given similar construction materials. The waterfall plot exhibited moderate temporal variation with persistent bright bands at the dominant WiFi channel locations and lower temporal variability between bursts compared to the urban outdoor case, reflecting the more stable propagation geometry of the enclosed office environment. These indoor spectral occupancy patterns are consistent with empirical spectrum occupancy measurements documenting elevated and temporally stable WiFi utilization in enclosed office and building environments \cite{Kuester2022,Chantaveerod2021}.

Table~\ref{tab:env_comparison} summarizes the key quantitative spectral observations across the three environments, enabling direct comparison of peak power, spectral dynamic range, channel occupancy, and temporal variability. The foliage environment is characterized by consistently low received power, minimal spectral structure, and low temporal variability, representing a severe attenuation scenario. The urban outdoor environment exhibits the highest dynamic range, the greatest number of identifiable emitters, and the highest temporal variability, driven by active WiFi traffic and multipath from surrounding structures. The indoor environment occupies an intermediate position, with strong channel occupancy but reduced dynamic range and moderate temporal variability relative to the urban case.

\begin{table}[H]
\centering
\caption{Cross-Environment Spectral Comparison at 2.45~GHz}
\begin{tabular}{|l|c|c|c|c|}
\hline
\textbf{Env.} & \textbf{Peak (dBFS)} & \textbf{DR (dB)} & \textbf{Ch.} & \textbf{Temp. Var.} \\
\hline
Foliage      & $-76$ to $-82$ & $\sim$6  & --  & Low  \\
\hline
Outdoor & $\sim$$-40$    & $\sim$30 & 3+  & High \\
\hline
Indoor       & $\sim$$-43$    & $\sim$20 & 2   & Mod. \\
\hline
\end{tabular}
\label{tab:env_comparison}
\end{table}

The geotagged SigMF dataset produced across these three environments constitutes a reproducible, environment-labeled corpus of real-world IQ recordings at 2.45~GHz. The spectral signatures observed within each environment confirm that the platform produces stable, reliable measurements, and that the environmental propagation differences are sufficiently pronounced to support downstream spectrum characterization, interference analysis, and environment-aware wireless research \cite{Romero2022,Nasser2021,Chandhok2021}.

\section{Discussion}

The experimental results demonstrate that a compact, cost-effective SDR platform can reliably capture high-fidelity, geotagged IQ data across operationally distinct environments while preserving the spectral fidelity and metadata richness required for reproducible propagation research. The three environments evaluated in this work produced propagation signatures that are not only qualitatively distinct but quantitatively separable, with peak power differences exceeding 35~dB between the foliage and urban outdoor cases and clear differences in spectral structure, channel occupancy, and temporal variability across all three settings. These results confirm that the platform is sufficiently sensitive and stable to capture environmentally meaningful propagation differences using accessible, commodity hardware, consistent with prior demonstrations of low-cost SDR instrumentation in rigorous field measurement contexts \cite{Wright2020,Flak2022,MolinaTenorio2021}.

The foliage results warrant particular attention in the context of ISM-band propagation modeling. The near-noise-floor power levels observed between $-76$ and $-82$~dBFS, combined with the absence of any identifiable WiFi channel structure, indicate that the dense deciduous canopy at the measurement site imposed attenuation on the order of 30~dB relative to the urban outdoor baseline, broadly consistent with reported values for dense vegetation at 2.4~GHz \cite{chateauvert2024safe,ma2024wave,zhang2023vegetation}. The slow temporal variation in the foliage waterfall, characterized by gradual horizontal striations rather than rapid burst activity, suggests that diffuse canopy scattering is the dominant propagation mechanism rather than specular reflection, reinforcing findings that standard exponential decay models consistently underestimate attenuation under realistic canopy density and moisture conditions \cite{BarriosUlloa2022,Hakim2022,CamaPinto2020}.

The urban outdoor results illustrate a qualitatively different regime dominated by multipath richness and emitter density. The 30~dB dynamic range, combined with at least three overlapping WiFi channel envelopes and frequent frequency-hopping transients consistent with Bluetooth and Zigbee activity, reflects the congested and temporally dynamic nature of the 2.45~GHz band in a dense campus environment \cite{shakya2022urban,samad2021campus,Polak2020,Sathya2021a}. The indoor results reveal a propagation environment not immediately predictable from path loss models alone: while outdoor-to-indoor penetration loss reduced absolute received power by an estimated 20--25~dB relative to the outdoor baseline \cite{ali-ooti-35g,DiagoMosquera2021}, the spectral baseline between WiFi channel peaks was simultaneously elevated by 8--10~dB, indicating that structural reflections create a persistent interference floor absent outdoors. This elevated baseline reduces the effective dynamic range available for identifying weak emitters within the building, a nuance with practical implications for indoor spectrum sensing and interference detection \cite{Kuester2022,Chantaveerod2021}.

The current field campaign represents a proof-of-concept evaluation of the platform rather than a comprehensive propagation study. Each environment was measured in a single session of approximately ten minutes at one location, which is sufficient to demonstrate the platform's operational capability and capture environment-dependent spectral differences, but does not support strong claims of statistical repeatability or generalized propagation modeling. The observed spectral signatures are indicative of the propagation conditions present at each site and are consistent with the broader literature, but broader conclusions would require multiple locations per environment, repeated sessions across different days and seasons, and longer capture durations to build the empirical foundation required for robust propagation characterization.

Beyond the environment-specific observations, these results collectively validate several design decisions embedded in the platform architecture. The use of a constant receiver gain setting across all three environments proved essential for enabling cross-environment power comparisons: had automatic gain control been applied, the absolute power differences between environments would have been obscured and the quantitative comparisons in Table~\ref{tab:env_comparison} would not have been possible. The 60-second SigMF segment structure balanced file management overhead against the need for sufficient temporal averaging, and the embedded GNSS metadata enabled unambiguous association of each segment with its physical capture location \cite{hilburn2017sigmf,Romero2022}, positioning the dataset for integration into radio environment mapping frameworks \cite{Reddy2022,Shrestha2023,Ivanov2022}. The absence of dropped samples, buffer underruns, or GNSS dropouts across the full campaign, combined with consistent write throughput in the 75--85~MB/s range, demonstrates that the Raspberry Pi~5 and HackRF One combination is capable of sustained professional-grade field operation, and that the barrier to generating high-quality, geotagged, environment-labeled IQ datasets is lower than prior work has implied \cite{Perotoni2021,tuta2023sdr,kandregula2024portable}.

\section{Conclusion}

This work presented the design, implementation, and field evaluation of a compact, battery-powered RF capture platform integrating a HackRF One software-defined radio, Raspberry Pi~5 single-board computer, embedded GNSS receiver, 2~TB solid-state drive, and a regulated battery power system into a fully autonomous, cost-effective unit. The platform sustains 20~Msps IQ capture while automatically embedding per-segment geolocation and temporal metadata into SigMF-compliant recordings \cite{hilburn2017sigmf}, enabling reproducible and context-aware spectrum measurements without the bulk, cost, or infrastructure dependencies of traditional RF instrumentation \cite{Wright2020,tuta2023sdr}.

Field evaluations conducted at 2.45~GHz across dense foliage, urban outdoor, and indoor office environments demonstrated that the platform produces stable, high-fidelity IQ datasets with clear and environment-consistent propagation signatures across all three settings. Foliage environments imposed severe attenuation, reducing received power to within 6~dB of the system noise floor and eliminating all identifiable spectral structure, consistent with canopy-induced losses on the order of 30~dB relative to the urban outdoor baseline and aligned with empirical vegetation attenuation findings in the literature \cite{chateauvert2024safe,ma2024wave,BarriosUlloa2022}. Urban outdoor environments produced spectrally rich recordings with a 30~dB dynamic range, multiple overlapping WiFi channel envelopes, and high temporal variability driven by active 802.11 traffic, Bluetooth and Zigbee frequency-hopping activity, and ISM-band co-channel interference \cite{shakya2022urban,Sathya2021a,Polak2020}. Indoor environments exhibited strong WiFi channel occupancy with an estimated 20--25~dB building entry loss relative to the outdoor baseline, alongside an elevated spectral interference floor of 8--10~dB attributable to structural multipath reflections, broadly consistent with outdoor-to-indoor penetration loss characterizations in the literature \cite{ali-ooti-35g,DiagoMosquera2021}. Across all sessions the platform sustained 75--85~MB/s write throughput with no dropped samples, no buffer underruns, and sub-second GNSS synchronization, confirming that commodity single-board-computer hardware is capable of professional-grade sustained field operation \cite{Flak2022,MolinaTenorio2021}.

The primary contributions of this work are a fully integrated, cost-effective RF capture platform with native SigMF compliance and embedded GNSS metadata, a structured three-environment field measurement methodology, and an initial quantitative characterization of 2.45~GHz propagation signatures across foliage, urban, and indoor settings from a proof-of-concept field campaign. The geotagged, environment-labeled IQ dataset produced by this campaign constitutes a reproducible corpus suitable for initial spectrum characterization, interference analysis, and radio environment mapping \cite{Romero2022,Reddy2022}. Together these contributions establish a practical and accessible foundation for scalable spectrum monitoring and environment-aware wireless research in operationally diverse settings.

\section{Future Work}

Expanding the measurement campaign to multiple locations per environment, repeated across seasons, would strengthen propagation characterization claims. The current dataset covers 2.45~GHz at one location per environment; extending to mid-band frequencies such as 3.5~GHz and 5.8~GHz would broaden applicability to 5G coexistence research \cite{garcia-penetration-35g,Munoz2023,Nguyen2021}. Seasonal campaigns are particularly valuable for foliage, where attenuation differs substantially between leafed and leafless conditions \cite{chateauvert2024safe,ma2024wave,Hakim2022}.

Multi-node deployment would enable spatial diversity analysis, distributed interference localization, and support for radio environment map construction \cite{Romero2022,Reddy2022,Shrestha2023}. Platform enhancements including a higher-resolution front-end (e.g., USRP B210 for 12-bit IQ), ruggedized enclosure, and UAV mounting for volumetric mapping \cite{Ivanov2022} represent natural extensions, while a compact real-time display would improve field situational awareness \cite{Chandhok2021,Nasser2021}.

\balance
\bibliographystyle{IEEEtran}
\bibliography{refs}

\end{document}